\documentstyle[epsfig]{aipproc}

\begin{document}
\title{The $\Lambda _0$ Polarization and the Recombination Mechanism\thanks{This
work was partially supported by Centro Latino Americano de F\'{\i}sica (CLAF).}}

\author{G. Herrera\dag, J. Magnin\ddag, Luis M. Monta\~no\dag and F.R.A. Sim\~ao\ddag}
\address{\dag CINVESTAV, Apdo postal 14-740 Mexico DF, Mexico \\ \ddag CBPF, Rua Dr. Xavier Sigaud 150, CEP 22290-180 - Urca RJ,Rio de Janeiro, Brazil}

\maketitle
\begin{abstract}
We use the recombination and the Thomas Precession Model to
obtain a prediction for the $\Lambda _0$ polarization in the $p+p\rightarrow \Lambda _0+X$ reaction. We study the effect of the recombination function on the $\Lambda_0$ polarization.
\end{abstract}

\section*{Introduction}
The unexpected discovery of large polarization in inclusive $\Lambda_0$
production by unpolarized protons has shown that important spin effects arise in the hadronization process. Several models have been proposed to explain hyperon polarization, being the Thomas Precession Model (TPM)\cite{degrand} one of the most extensively used to describe polarization in a variety of reactions.

In order to obtain the $\Lambda_0$ polarization, we first calculate the momentum fraction of the recombining $s$-quark in the proton sea using a recombination model \cite{herrera}. We use two different forms for the recombination function to see their influence on the predicted $\Lambda_0$ polarization.\\

\section*{\label{Dos}The $\Lambda _0$ polarization in the TPM}

In $pp$ collisions the recombining $s$-quark resides in the sea of the proton and carries a very small fraction $x_s\simeq 0.1$ of the proton momentum. When
the $s$-quark recombines to form a $\Lambda _0$, it becomes a valence quark
and must carry a large fraction (of the order of $\frac 13$) of $
\Lambda _0$'s momentum. Then one expects a large increase in the longitudinal momentum of the $s$-quark as it passes from the proton to the $\Lambda _0$,
\begin{equation}
\Delta p\simeq (\frac 13x_F-x_s)p=\left( \frac 13-\xi \right) x_Fp,
\label{uno}
\end{equation}
where $p$ is the proton's momentum, $\xi =x_s/x_F$ and $x_Fp=p_\Lambda $ is the momentum of the $\Lambda _0$ with $x_F$ the Feynman $x$.

Since the $s$-quark carries transverse momentum, on the average $p_T(s/p)\sim p_T(s/\Lambda )\sim \frac 12p_{T\Lambda }$, its velocity vector is not parallel to the change in momentum induced by recombination and it must feel the effect of Thomas precession. Consequently, the $\Lambda_0$ is produced with net polarization perpendicular to the plane of the reaction.

According to the TPM the $\Lambda_0$ polarization in the reaction $p+p\rightarrow\Lambda_0 +X$ is given by\cite{degrand}
\begin{equation}
P(p\rightarrow \Lambda )=-\frac 3{M^2\Delta x}\frac{\left( 1-3\xi \right) }{%
\left( \frac{1+3\xi }2\right) ^2}p_{T\Lambda },  
\label{tres}
\end{equation}
where $M^2=\left[ \frac{m_D^2+p_{TD}^2}{1-\xi }+\frac{m_s^2+p_{Ts}^2}\xi -m_\Lambda^2-p_{T\Lambda }^2\right]$ and $\xi =\frac 13\left(1-x_F\right) +0.1x_F$ as was assumed in ref. \cite{degrand}. $\Delta x=0.5$ GeV is a characteristic recombination scale and $m_D$, $p_{TD}$, $m_s$, $p_{Ts}$, $m_\Lambda $ and $p_{T\Lambda }$ are respectively the masses and transverse momentum of the diquark, the $s$-quark and the $\Lambda_0 $.

\section*{The $\xi \left( x_F\right) $ parametrization in the Recombination Model}

We use the recombination model proposed in ref. \cite{das-hwa}, which has been extended to take into account baryon production \cite{ranft}, to obtain a parametrization for $\xi$ as a function of $x_F$\cite{herrera}. 
The inclusive $x_F$ distribution for $\Lambda_0$'s in $p p$ collisions is
\begin{equation}
\frac{d\sigma }{dx_F}=\int \frac{dx_u}{x_u}\frac{dx_d}{x_d}\frac{dx_s}{x_s}
F\left( x_u,x_d,x_s\right) R\left(x_F,x_u,x_d,x_s\right) ,  
\label{cinco}
\end{equation}
where $F(x_{u},x_{d},x_{s})$ and $R(x_{F},x_{u},x_{d},x_{s})$ are the three quark distribution and recombination functions respectively.

For the three quark distribution function we use the factorized form
\begin{equation}
F\left( x_u,x_d,x_s\right) =\beta F_{u,val}\left( x_u\right) F_{d,val}\left(
x_d\right) F_{s,sea}\left( x_s\right) \left( 1-x_u-x_d-x_s\right) ^\gamma 
\label{seis}
\end{equation}
with $\gamma =-0.3$ as has been proposed in ref. \cite{ranft} and $\beta
=0.75$. We used the Field and Feynman \cite{feynman} parametrizations for
the single quark distribution.

In order to see how the shape of the recombination function affects the prediction for the $\Lambda_0$ polarization, we use two different forms for $R(x_u,x_d,x_s)$:
\begin{equation}
R_1\left( x_u,x_d,x_s\right) =\kappa _1 \frac{x_ux_dx_s}{\left( x_F\right) ^3}%
\delta \left( \frac{x_u+x_d+x_s}{x_F}-1\right) 
\label{siete}
\end{equation}
as in ref. \cite{ranft} and
\begin{equation}
R_2\left( x_u,x_d,x_s\right) =\kappa _2 \left( \frac{x_ux_d}{x_F^2}\right)^{a}
\left( \frac{x_s}{x_F}\right) ^{b}\delta \left( \frac{x_u+x_d+x_s}{x_F%
}-1\right) ,  \label{ocho}
\end{equation}
which is inspired in the three valons recombination model proposed by R.C. Hwa \cite{hwa}. In $R_2$, unlike $R_1$, the light quarks are considered with
different weight than the more massive $s$ quark introducing two distinct
exponents $a$ and $b$. Indeed, in the recombination model proposed in ref.
\cite{hwa}, a recombination function for hyperons is derived and a ratio 
${\frac ab} = {\frac 23}$ is used. We choose $a = 1$, $b = {\frac 32}$ by fitting experimental data. $\kappa _1 $ and $\kappa _2 $ are normalization constants. 

The probability for $\Lambda _0$ production at $x_F$ with an $s-quark$ from
the sea of the proton at momentum fraction $x_s$ is
\begin{equation}
\frac{d\sigma _i}{dx_sdx_F}=\int \frac{dx_u}{x_u}\frac{dx_d}{x_d}\frac
1{x_s}F\left( x_u,x_d,x_s\right) R_i\left(x_F,x_u,x_d,x_s\right)
\label{nueve}
\end{equation}
with $i=1,2$. The average value of $x_s$ is therefore \cite{herrera}
\begin{equation}
\left\langle x_s\right\rangle _i=\left[\int dx_sx_s\frac{d\sigma_i}{dx_sdx_F}\right]/ \frac{d\sigma _i}{dx_F}.  
\label{diez}
\end{equation}

\begin{figure}[t!] 
\centerline{\epsfig{file=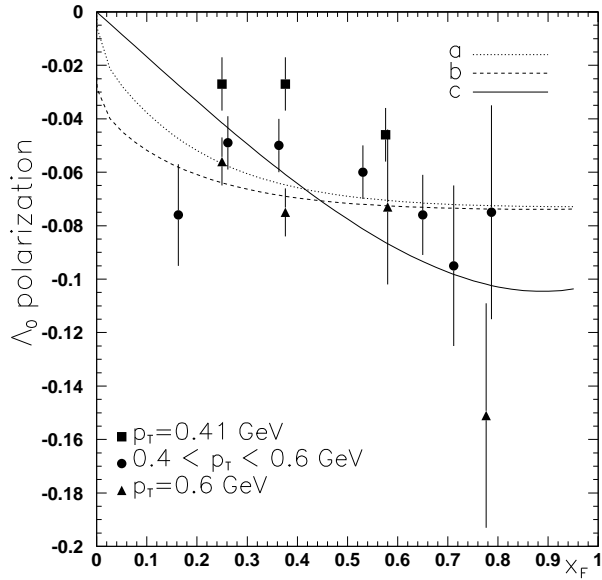,height=3.5in,width=3.5in}}
\vspace{5pt}
\caption{$\Lambda _0$ polarization at $p_T=0.5 GeV/c $ obtained with $\xi (x_F)$ determined with the recombination functions $R_1$ (a), and $R_2$ (b). (c) is the polarization prediction of ref. [1]. Experimental data are taken from refs. [1] and [7].}
\label{fig1}
\end{figure}

We have taken $m_D=\frac 23$ GeV, $m_s=\frac 12$ GeV and ${\left\langle p_{T}^2\right\rangle}_{s,D} = \frac 14{p_{T\Lambda}^2}+\left\langle k_{T}^2\right\rangle $ with $\left\langle k_{T}^2\right\rangle =0.25$ GeV $^2$ \cite{degrand}. The figure \ref{fig1} shows the $\Lambda _0$ polarization for the three different parametrizations of $\xi(x_F)$ at $p_T=0.5$ GeV/c.

\section*{Conclusions}

The two forms for $\xi$ obtained with the two different recombination functions of eqs. \ref{siete} and \ref{ocho} are very similar in shape for large $x_F$. For small $x_F$ however, the difference grows slightly and $\xi _1\left( x_F=0\right) =\frac 13$ while $\xi _2\left( x_F=0\right) \neq\frac 13$.

The parametrizations for $\xi \left( x_F\right) $ obtained from the recombination model are different to the simple form proposed in ref. \cite
{degrand}. Our calculation of $\xi \left( x_F\right) $ shows that, for $x_F
\rightarrow 1 $, $\xi \left( x_F\right) \rightarrow 0.15 $ approximately for
both recombination functions. This is consistent with our actual knowledge of the sea structure functions in the proton.

We have seen that for small $p_{T\Lambda }$ our fit gives a good  description of ex\-pe\-ri\-men\-tal data. This is reasonable since recombination models work better for small $p_{T}$.

Within the precision of experimental data\cite{das-hwa}[7], it would be hard to decide which recombination function better describe $\Lambda _0$'s production. A more accurate measurement of polarization at low $p_T$ and low $x_F$ can help to clarify the right form of the recombination function. It is interesting to note that, although the shape of the recombination function is not important for cross section calculations, it does make a difference when applied to polarization. In this sense, polarization measurements can help to understand the underlying mechanisms in hadroproduction.

\section*{Acknowledgments}

We would like to thank the organizers for financial support to attend the I Simposium Latino Americano de F\'{\i}sica de Altas Energ\'{\i}as.

\end{document}